\documentclass{aa}  

\usepackage{graphicx}
\usepackage{txfonts}
\begin{document} 

   \title{The GTC exoplanet transit spectroscopy survey VII}

   \subtitle{An optical transmission spectrum of WASP-48b}

   \author{F. Murgas
          \inst{1}\fnmsep\inst{2}\fnmsep\inst{3}\fnmsep\inst{4}
          \and
          E. Pall\'{e}\inst{1}\fnmsep\inst{2}
          \and
          H. Parviainen\inst{1}\fnmsep\inst{2}
          \and
          G. Chen\inst{1}\fnmsep\inst{2}\fnmsep\inst{7}
          \and
          L. Nortmann\inst{1}\fnmsep\inst{2}
          \and
          G. Nowak\inst{1}\fnmsep\inst{2}
          \and
          A. Cabrera-Lavers\inst{1}\fnmsep\inst{5}
          \and
          N. Iro\inst{6}
          }

   \institute{Instituto de Astrof\'isica de Canarias (IAC), E-38205 La Laguna, Tenerife, Spain
         \and
             Departamento de Astrof\'isica, Universidad de La Laguna (ULL), E-38206 La Laguna, Tenerife, Spain
         \and
         Univ. Grenoble Alpes, IPAG, F-38000 Grenoble, France    
         \and
         CNRS, IPAG, F-38000 Grenoble, France
         \and
         Gran Telescopio Canarias (GTC), E-38712, Bre\~{n}a Baja, La Palma, Spain
         \and
         Theoretical Meteorology group, Klimacampus, University of Hamburg, Grindelberg 5, 20144 Hamburg, Germany
         \and
         Key Laboratory of Planetary Sciences, Purple Mountain Observatory, Chinese Academy of Sciences, Nanjing 210008, China
   }

   \date{Received 04 April 2017/ Accepted 22 June 2017}

 
  \abstract
   {Transiting planets offer an excellent opportunity for characterizing the atmospheres of extrasolar planets under very different conditions from those found in our solar system.}
   {We are currently carrying out a ground-based survey to obtain the transmission spectra of several extrasolar planets using the 10 m Gran Telescopio Canarias. In this paper we investigate the extrasolar planet WASP-48b, a hot Jupiter orbiting around an F-type star with a period of 2.14 days.}
   {We obtained long-slit optical spectroscopy of one transit of WASP-48b with the Optical System for Imaging and low-Intermediate-Resolution Integrated Spectroscopy (OSIRIS) spectrograph. We integrated the spectrum of WASP-48 and one reference star in several channels with different wavelength ranges, creating numerous color light curves of the transit. We fit analytic transit curves to the data taking into account the systematic effects present in the time series in an effort to measure the change of the planet-to-star radius ratio ($R_\mathrm{p}/R_\mathrm{s}$) across wavelength. The change in transit depth can be compared with atmosphere models to infer the presence of particular atomic or molecular compounds in the atmosphere of WASP-48b.}
   {After removing the transit model and systematic trends to the curves we reached precisions between 261 ppm and 455-755 ppm for the white and spectroscopic light curves, respectively. We obtained $R_\mathrm{p}/R_\mathrm{s}$ uncertainty values between $0.8 \times 10^{-3}$ and $1.5\times 10^{-3}$ for all the curves analyzed in this work. The measured transit depth for the curves made by integrating the wavelength range between 530 nm and 905 nm is in agreement with previous studies. We report a relatively flat transmission spectrum for WASP-48b with no statistical significant detection of atmospheric species, although the theoretical models that fit the data more closely include of TiO and VO.}
   {}


   \maketitle
%

\section{Introduction}
With more than 2700 confirmed transiting extrasolar planets\footnote{http://exoplanet.eu/catalog/} and many unconfirmed candidates, the characterization of their atmospheres through transmission spectroscopy has become an active subfield in exoplanetary studies. Although the first detection of elements using transmission spectroscopy were made from space (e.g., \citealp{Charbonneau2002}, \citealp{VidalMadjar2004}, \citealp{Lecavelier2008}, \citealp{Desert2009}), ground-based telescopes and instruments are also able to achieve the high precision measurements required by this technique. Indeed, after the first space-based discoveries of atoms and molecules in extrasolar planets, many ground-based observations followed (e.g., \citealp{Snellen2008}, \citealp{Redfield2008}, \citealp{Bean2011}, \citealp{Sing2012}, \citealp{Bean2013}) and at present there are several efforts to continue the characterization of exoplanetary atmospheres from the ground.

We are currently carrying out a transmission spectroscopy survey to characterize the atmospheres of transiting planets using long-slit spectroscopy with the 10.4 m telescope Gran Telescopio Canarias (GTC). The principle of transmission spectroscopy is to measure the change in the planetary radius at different wavelengths during a transit event, and to use this change to infer the presence of particular components of the exoplanet atmosphere. Because of their relatively big atmospheric scale heights, hot Jupiters are excellent targets for atmospheric detection from the ground and they are the main targets of our survey (e.g., \citealp{Murgas2014}, \citealp{Parviainen2016}, \citealp{Palle2016}, \citealp{Nortmann2016}, \citealp{Chen2017}).

The hot Jupiter WASP-48b was discovered by \citet{Enoch2011} as part of the Wide Angle Search for Planets (WASP, \citealp{Pollacco2006}). Its host star is a slightly evolved F star with an apparent magnitude of $V = 11.72 \pm 0.14$ mag, a mass of $M = 1.19 \pm 0.05$ $M_\odot$, a stellar radius of $R = 1.75 \pm 0.09$ $R_\odot$, and an effective temperature of $T_{eff}= 6000 \pm 150$ K. \citet{Enoch2011} measured a planetary mass of $M = 0.98 \pm 0.09$ $M_{Jup}$, a radius of $R=1.67 \pm 0.10$ $R_{Jup}$, and an orbital period of 2.14 days. \citet{Sada2012} measured a planet-to-star radius ratio in the $J$ band of $R_\mathrm{p}/R_\mathrm{s} = 0.0988^{+0.0051}_{-0.0049}$, which agrees with the value reported in the discovery paper by \citet{Enoch2011} ($R_\mathrm{p}/R_\mathrm{s} = 0.0980 \pm 0.001$). \citet{ORourke2014} used \textit{Spitzer} to measure the secondary transit of WASP-48b in the $H$, $K_s$, 3.6 $\mu$m, and 4.5 $\mu$m bands. For the dayside emission spectra they fitted a blackbody model with an effective temperature of $T_{eff} = 2158 \pm 100$ K. Using the models of \citet{Fortney2008} and \citet{Burrows2008} to fit their data, they deduced that WASP-48b has a weak or absent temperature inversion. \citet{Ciceri2015} presents a follow up of this system using multicolor broadband photometry; this work updated the stellar parameters of WASP-48 ($M = 1.062 \pm 0.074$ $M_\odot$, $R = 1.519 \pm 0.007$ $R_\odot$), measured a smaller planetary radius ($R=1.396 \pm 0.051$ $R_{Jup}$), and found a flat transmission spectrum using broadband filters. \citet{Turner2016} presents ground-based near-UV data of several transiting systems including WASP-48b, and found a $R_\mathrm{p}/R_\mathrm{s}$ for the $U$ band of $0.0916 \pm 0.0017$, a shallower transit depth than the one found by \citet{Enoch2011}.

This paper is organized as follows. In \S \ref{sec:obs} and \S \ref{sec:datred} we describe the observations and data reduction, in \S \ref{sec:fitting} we describe the light curve fitting process, and in \S \ref{sec:results} we show the transmission spectrum of WASP-48b. Finally, in \S \ref{sec:discussion} we present a discussion of the results and in \S \ref{sec:conclusions} the conclusions of this paper.

\section{Observations}
\label{sec:obs}
The GTC is a 10.4 m telescope located at Observatorio Roque de los Muchachos in La Palma, Spain. For the observations, the Optical System for Imaging and low-Resolution-Integrated Spectroscopy (OSIRIS, \citealp{Cepa2000}) in its long-slit spectroscopic mode was chosen.

The data were taken on July 17, 2014; the binning was set to $2\times 2$; the readout speed of OSIRIS was chosen to be 200 KHz with a gain of 0.95 e$^{-}$/ADU; and readout noise of 4.5 e$^{-}$. The observations were made using the R1000R grism (spectral coverage of $\lambda \sim 520-1040$ nm) and with a custom built slit with a width of 40 arcsec. The use of a wide slit minimizes the systematic effects that may appear due to flux losses caused by seeing variations during the observations. The exposure time was set to 15 seconds, thus assuring a high signal-to-noise ratio below the saturation level of the detector. Close to 50 minutes of data before and after the transit were collected; the total time of data acquisition was $5$ hours, which translated into 552 science images. During the observations, the airmass varied from 1.32 to 1.22 (minimum airmass of 1.19, maximum telescope elevation angle was 63.2$^\circ$). 

\begin{figure}
    \centering
    \includegraphics[width=\hsize]{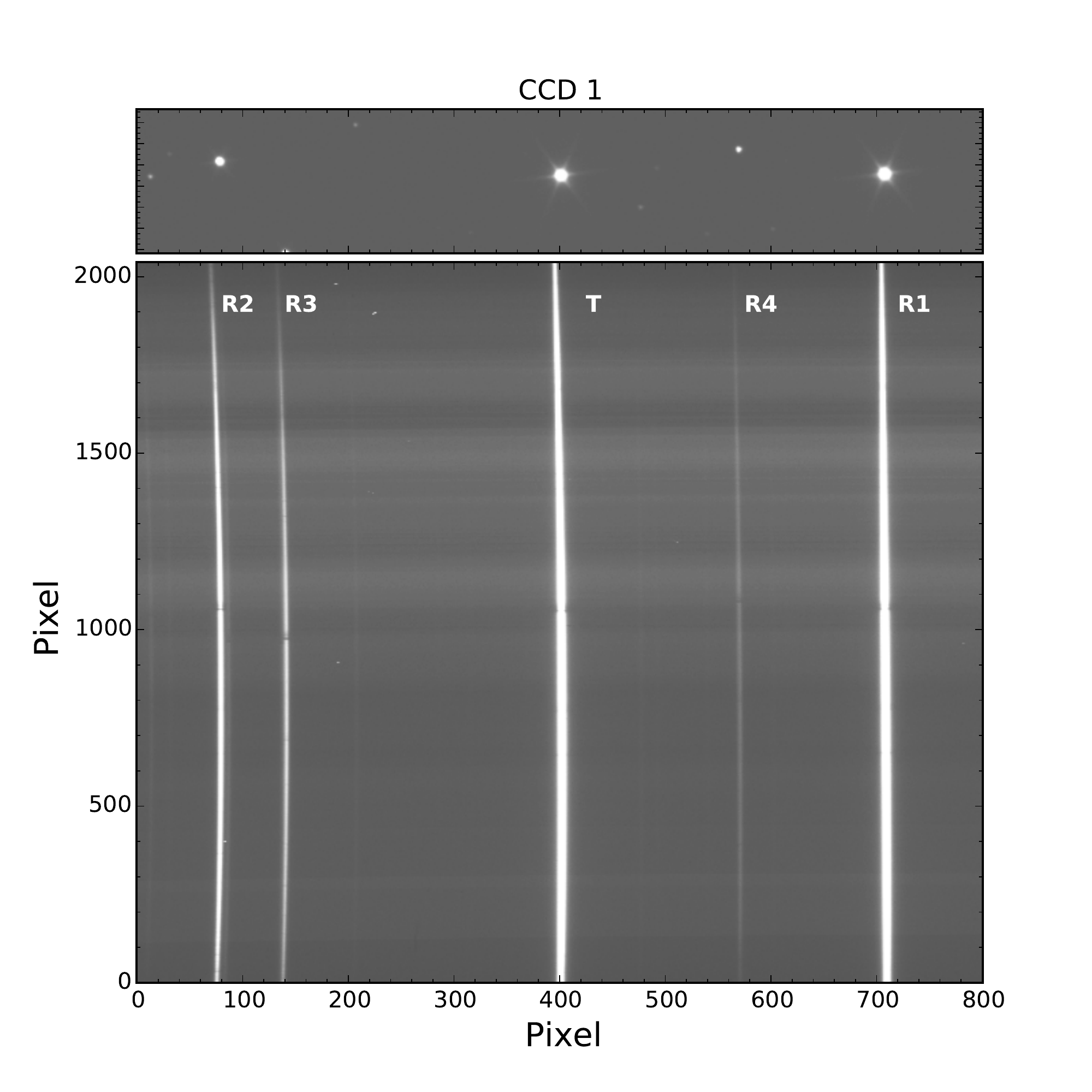}
    \caption{GTC OSIRIS image through the slit (top) and calibrated science image (bottom). In this image, the target is the star labeled T and the selected reference star is labeled R1. Other reference stars in the field (R2, R3, and R4) are marked, but were not used in the data analysis due to their faint magnitude.}
    \label{fig:CCD}
  \end{figure}

During the transit of WASP-48b, OSIRIS was able to obtain spectra for the target and four reference stars (Figure \ref{fig:CCD}) simultaneously. Because of its low brightness, three of the stars were not suitable to be used as reference and were discarded from the analysis. All the results presented here were produced using the flux of the star R1 as reference (see Table \ref{table:StarDescr}).

The universal time of data acquisition was determined using the recorded headers of the images. The opening and closing time of the shutter was used to compute the time of mid exposure. Then, using the code written by \citet{Eastman2010}\footnote{http://astroutils.astronomy.ohio-state.edu/time/}, the mid exposure time was converted to the Barycentric Julian Date in Barycentric Dynamical Time (BJD\textunderscore TDB). All the results present here were produced using BJD as the time standard.

  \begin{table*}
    \caption{Observed stars.}
    \label{table:StarDescr}  
    \centering                          
    \begin{tabular}{ccccccc}        
    \hline\hline                 
    Star & 2MASS name & RA (FK5) & DEC (FK5) & B (mag)  & V (mag) & K (mag)\\    
    \hline                
    WASP-48 & J19243895+5528233 & 19$^h$ 24$^m$ 38.984$^s$ & +55$^\circ$ 28$^m$ 23.39$^s$ & 12.31 & 11.65 & 10.37 \\
    R1 & J19243079+5527485 & 19$^h$ 24$^m$ 30.787$^s$ & +55$^\circ$ 27$^m$ 48.52$^s$ & 11.99 & 11.44 & 10.45 \\
    \hline                                   
    \end{tabular}
  \end{table*}

\section{Data reduction}
\label{sec:datred}
The raw data were calibrated using standard procedures. With the calibration images provided by the GTC team, we proceeded to create an average bias using 32 images. We requested a large number of flats in order to have a good estimation of the pixel sensitivy of OSIRIS; the flat field calibration set was composed of 100 images that were combined after subtracting the average bias. The master flat was then normalized by fitting (including rejection of deviant pixels) a high degree Chebyshev polynomial to the flux across the dispersion axis.

\begin{figure}
   \centering
   \includegraphics[width=\hsize]{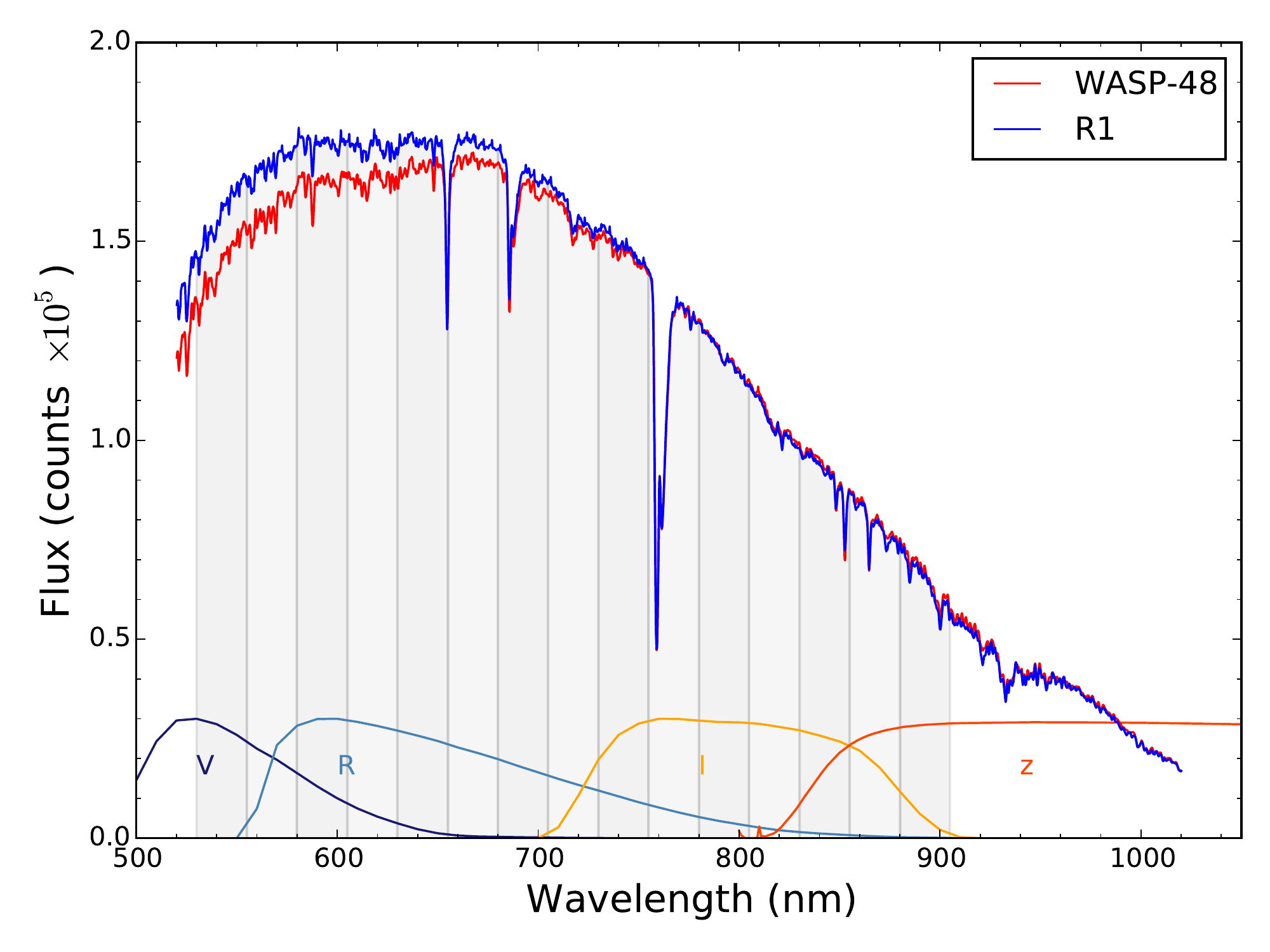}
   \caption{Extracted R1000R grism spectrum of WASP-48 (red) and its reference star (blue). The spectra are not corrected for instrumental response or flux calibrated. The shaded gray areas indicate the custom 25 nm passbands used to create the spectroscopic light curves. The broadband filters $V$, $R$, $I$, and $z-sloan$ are also plotted in an arbitrary scale to show the wavelength coverage of the spectrum.}
   \label{fig:Spectra}
 \end{figure}

Following \citet{Murgas2014}, the extraction and wavelength calibration of the spectra was done using a customized PyRAF script that calls different IRAF\footnote{IRAF is distributed by the National Optical Astronomy Observatory, which is operated by the Association of Universities for Research in Astronomy (AURA) under a cooperative agreement with the National Science Foundation.} tasks that work with long-slit spectroscopic data. A fixed aperture size of 20 pixels in width (5.08 arcsec) was used to extract the spectra. This aperture width was chosen after determining which one delivered the lowest scatter in the out-of-transit points in the white light curve (wavelength range integrated between 530-905 nm). In order to correct for drifts along the dispersion axis, the cross-correlation between the first image and the rest of the time series for each star was computed with IRAF's task FXCOR. This process was repeated between the target and the reference star. Additionally, a Gaussian function was fitted to the spectral profile of each star in the time series to monitor drifts perpendicular to the dispersion axis. The computed full width at half maximum (FWHM) of the spatial profile for each star was used in the fitting procedure as a proxy of the seeing variations observed during the transit. An example of the extracted spectrum and wavelength coverage of GTC R1000R grism can be seen in Figure \ref{fig:Spectra}.

 \begin{figure}
    \centering
    \includegraphics[width=\hsize]{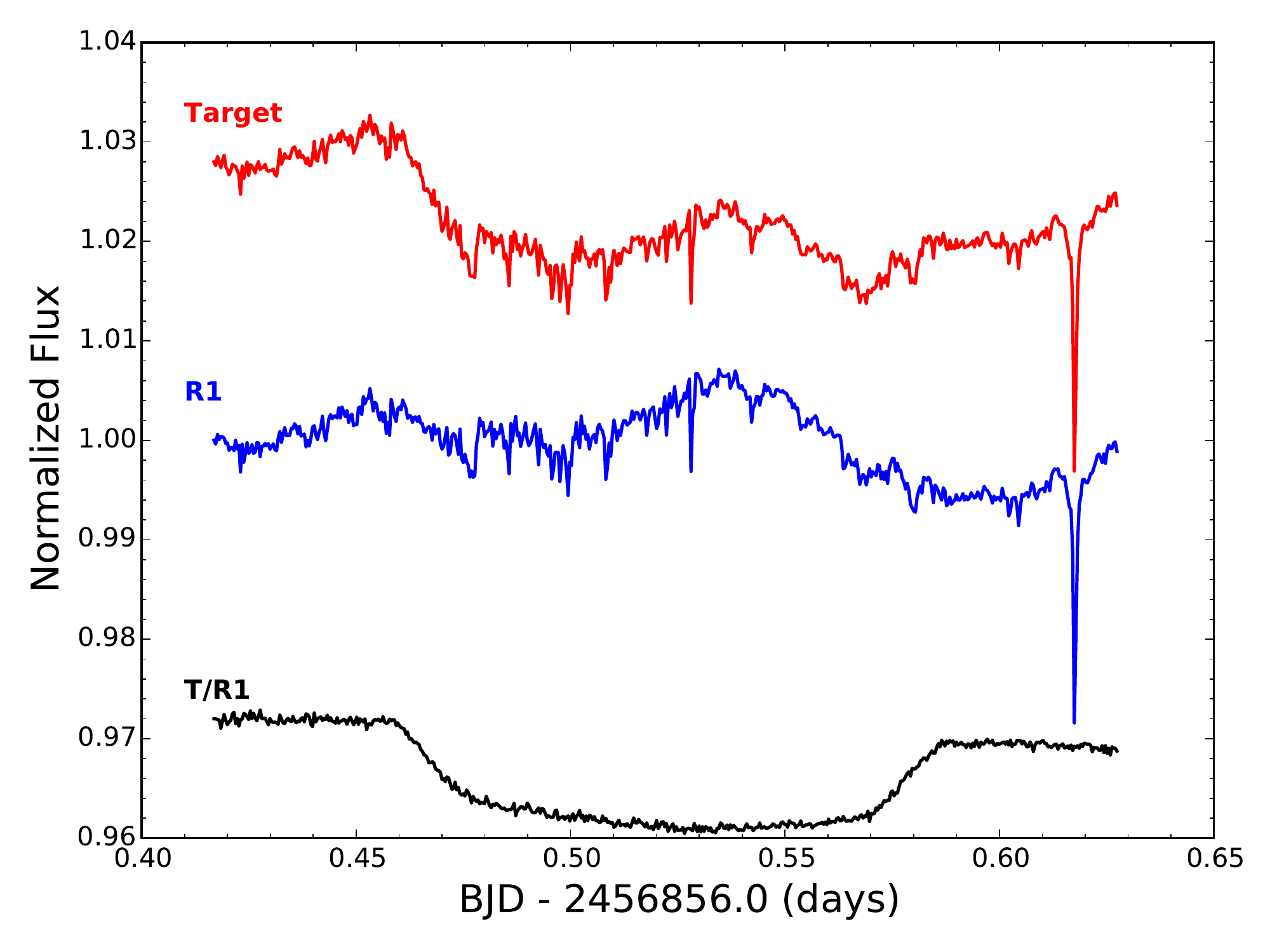}
    \caption{Individual light curves of the target and reference star (R1). In addition, the flux ratio between the target and the reference star is plotted for the white light curve. The fluxes are normalized to the first data point of the time series and with an arbitrary offset in the y-axis.}
    \label{fig:FluxStars}
 \end{figure}

\section{Light curve fitting}
\label{sec:fitting}
To create the light curves, the spectra of the target and reference star were integrated in different wavelength ranges and with different bin sizes. Then, the integrated flux of the target was divided by the flux of the reference star.

The curves created and analyzed in this work are as follows: 1 white light curve (flux integrated between 530 nm and 905 nm), 15 curves using a bin size of 25 nm (between 530 nm and 905 nm), and 6 curves made with a bin size of 10 nm (covering the spectral region 564.5 nm - 624.5 nm).

Figure \ref{fig:FluxStars} presents the flux of WASP-48, the reference star R1, and the flux ratio between the target and R1 for the white light curve. The plot shows clearly that variations in the flux of the target and reference star were present during the observations, but they are mostly corrected by the differential photometry. However, there are some residual systematic effects still present in the light curve which need to be taken into account in the fitting procedure in order to have a robust error estimation.

\subsection{Model selection}
As in \citet{Nortmann2016}, the Bayesian Information Criterion (BIC) was used to select the best light curve model. The BIC penalizes a high $\chi^2$ value and/or the number of free parameters used in the fitting procedure; the model that presents the lowest BIC is the one selected.

The BIC values for three models were computed for 16 light curves produced using a bin size of 25 nm of width covering a wavelength range between 530 nm and 905 nm. The following models were tested:

\begin{equation}
  \mathcal{M}_1 = \mathcal{T}(p)(a_0 + b_1t + c_0Fw),
  \label{Eq:Model1}
\end{equation}

\begin{equation}
  \mathcal{M}_2 = \mathcal{T}(p)(a_0 + b_1t + c_0Fw + c_1Fw^2),
  \label{Eq:Model2}
\end{equation}

\begin{equation}
  \mathcal{M}_3 = \mathcal{T}(p)(a_0 + b_1t + b_2t^2 + c_0Fw),
  \label{Eq:Model3}
\end{equation}
where $\mathcal{T}(p)$ is the transit model and $p$ are the transit dependent parameters, $t$ is the time of the observations, and $Fw$ is the full width at half maximum (FWHM) of the spectrum profile. All the tested models present a synthetic transit model with a time dependency to reproduce a second-order color effect not corrected by the differential photometry, and a FWHM dependent polynomial to model the systematic effects produced by seeing variations. The FWHM correction was introduced because seeing variations affected the number of pixels in which the major part of the stellar flux is contained, and this pixel-dependent effect will act as a systematic correlated with FWHM (see \citealp{Nortmann2016}, \citealp{Chen2017}). With a slit 40 arcsec in width, the flux losses produced by not being able to measure the flux contained in the wings of the spectral profile are negligible. Other possible sources for the systematic effects were tested (e.g., position of the stars, telescope rotation angle as in \citealp{Nortmann2016}, etc.), but no strong correlation was found and they were discarded as parameters for the tested models. 

  \begin{figure}
   \centering
   \includegraphics[width=\hsize]{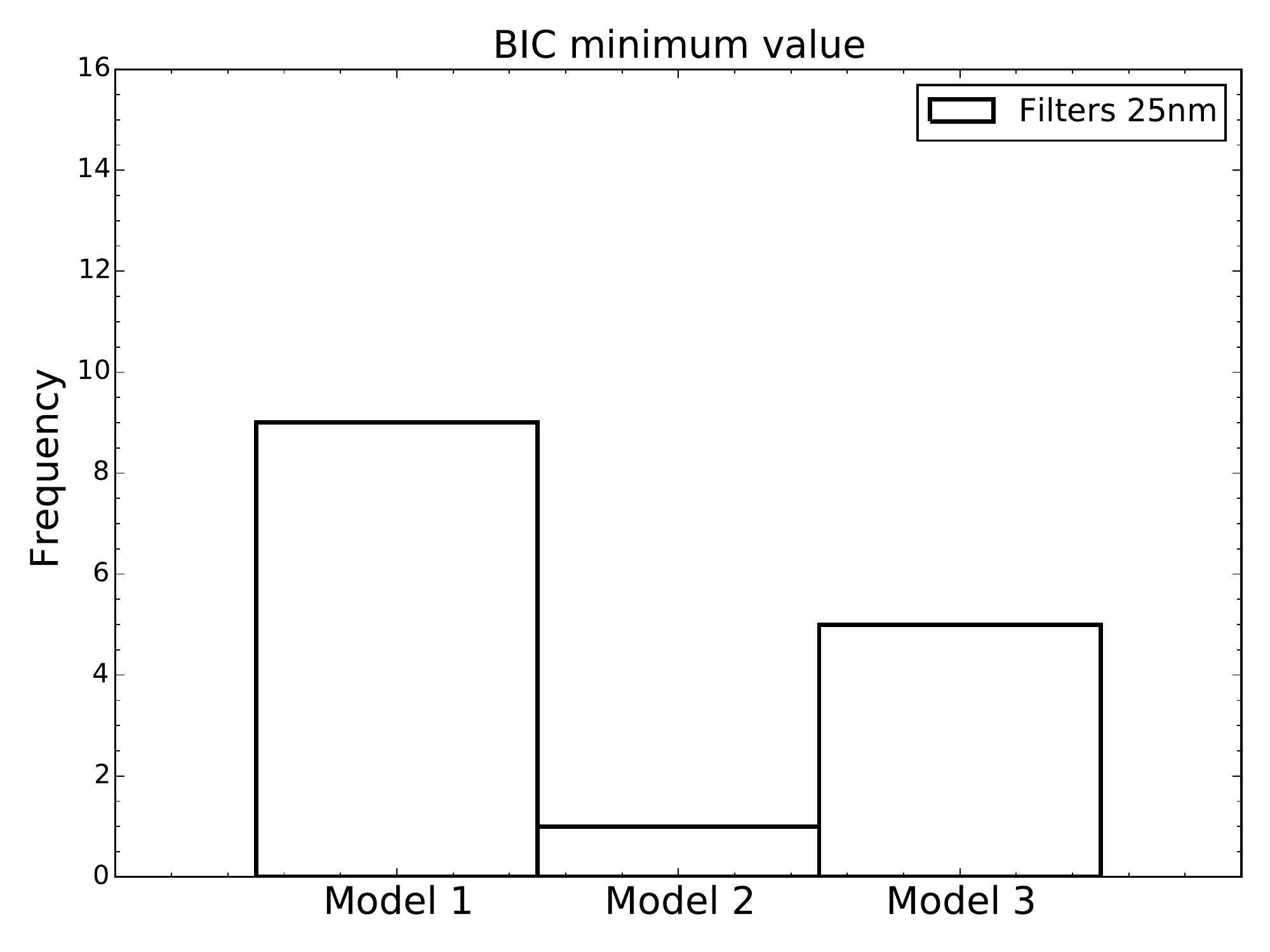}
   \caption{Frequency of the minimum BIC values for each proposed model. The model that presents the lowest BIC is the one selected. The BIC values were computed using curves created by integrating the flux of the stars using bins of 25 nm in width.}
   \label{fig:ModelSelec}
  \end{figure}

As shown in Figure \ref{fig:ModelSelec} no model was selected unanimously; for each curve created there were some bins that preferred a model (i.e., lowest BIC value) that did not get the majority of preferences. Since a unique model that presented the lowest BIC value for all the curves evaluated in this test could not be found, the function that provided the lowest BIC values at the highest rate was selected; in this case Model 1. Using a model with a high number of parameters increases the risk of overfitting the data; however, for most of the bins where the BIC was evaluated the numerical difference between Models 1, 2, and 3 was not significant.

\subsection{Fitting procedure}
In order to produce the synthetic light curves, we used the curve generator code \textit{PyTransit}\footnote{https://github.com/hpparvi/PyTransit} (\citealp{Hannu2015}). This code presents optimized Python routines that implement the \citet{Gimenez2006} transit models. 

For the transit curves, a quadratic limb darkening law was adopted. The limb darkening coefficients were modeled using the Python Limb Darkening Tool Kit (PyLDTK\footnote{https://github.com/hpparvi/ldtk}; \citealp{Parviainen2015}), this package uses \citet{Husser2013} stellar libraries to compute the coefficients using custom passbands. The stellar parameters used as input for PyLDTK were taken from the discovery paper of WASP-48b by \citet{Enoch2011}: $T_{eff}= 6000 \pm 150$ K, $\log g = 4.50 \pm 0.15$, and [Fe/H] $= -0.12 \pm 0.12$.

For each curve, we set the following transit parameters as free: the planet-to-star radius ratio $R_\mathrm{p}/R_\mathrm{s}$, the quadratic limb darkening coefficients $u_1$ and $u_2$, the central time of transit $T_c$, the orbital semi-major axis over stellar radius $a/R_s$, and the orbital inclination $i$. The orbital eccentricity and period were fixed to 0 and 2.14363544 days (\citealp{Ciceri2015}), respectively. In addition to the free transit parameters, we also set as free the coefficients used to model the systematic trends present in the light curves.

In order to estimate robustly the values of the modeled parameters, a Bayesian fitting process was used. The procedure described here is similar to the one presented in \citet{Parviainen2016}. A likelihood function was evaluated iteratively using the Python MCMC suite \textit{emcee} (\citealp{ForemanMackey2013}). The likelihood function was given by:
\begin{equation}
\ln \mathcal{L} = \ln \mathcal{L}_{curve} + \ln \mathcal{L}_{LD},
\end{equation}
where $\mathcal{L}_{curve}$ is the likelihood of the transit light curve and $\mathcal{L}_{LD}$ is the likelihood of the limb darkening coefficients. The likelihood of the curves is given by the comparison between the data and the transit model plus the systematic trends (see Eq. \ref{Eq:Model1}), and it was computed together with red noise estimation using Gaussian processes with a simple exponential kernel (the \textit{George} package for Python, \citealp{hodlr}). The likelihood of the limb darkening coefficients was computed using PyLDTK and it provides a more robust way of establishing the coefficient values instead of the traditional approach of using tables (\citet{Parviainen2016}). Only uniform priors were used for the parameters fitted in this work (see Table \ref{table:priors} for the transit parameters priors).

  \begin{table}
    \caption{Transit parameters uniform prior ranges.}
    \label{table:priors}
    \centering
    \begin{tabular}{cc}
    	\hline\hline 
    	Transit Parameter & Uniform prior range \\
    	\hline 
    	$R_\mathrm{p}/R_\mathrm{s}$ & $[0.03,0.15]$ \\
	$u_1$ & $[-1.0,1.0]$ \\
	$u_2$ & $[-1.0,1.0]$ \\
	$T_c-2456856$ [days] & $[0.512256,0.533096]$ \\
	$a/R_s$ & $[3.2,5.8]$ \\
	$i$ [deg] & $[75,90]$ \\
	\hline                                   
    \end{tabular}
 \end{table}

The fitting procedure was split in three stages. First, we performed a global optimization of $\mathcal{F}=\ln \mathcal{L} + \ln Priors$, i.e., the likelihood weighted by the priors, using \textit{PyDE}\footnote{https://github.com/hpparvi/pyde}. The second part of the process consisted in running a small MCMC (60 independent chains, 10000 iterations) as a burn-in period, using the global optimization results as a seed. Finally, we used the parameter vector that delivered the highest $\mathcal{F}$ value from stage 2 as a starting point to generate 250 independent chains and ran the MCMC procedure for 11500 iterations using \textit{emcee}. After this process was over, the autocorrelation length in the chains was calculated in order to avoid strongly correlated parameter values in the posterior probability distribution. The final values and uncertainties were obtained by computing the percentiles of the distributions to get the median and $1\sigma$ limits for each parameter.

Following \citet{Parviainen2016}, the two red noise parameters used to describe the exponential kernel (the amplitude and time scale), where computed for the white light curve and fixed for the rest of the curves (i.e., the 25 nm and 10 nm bins).  

\section{Results}
\label{sec:results}

\subsection{White light curve}
\begin{figure}
   \centering
   \includegraphics[width=\hsize]{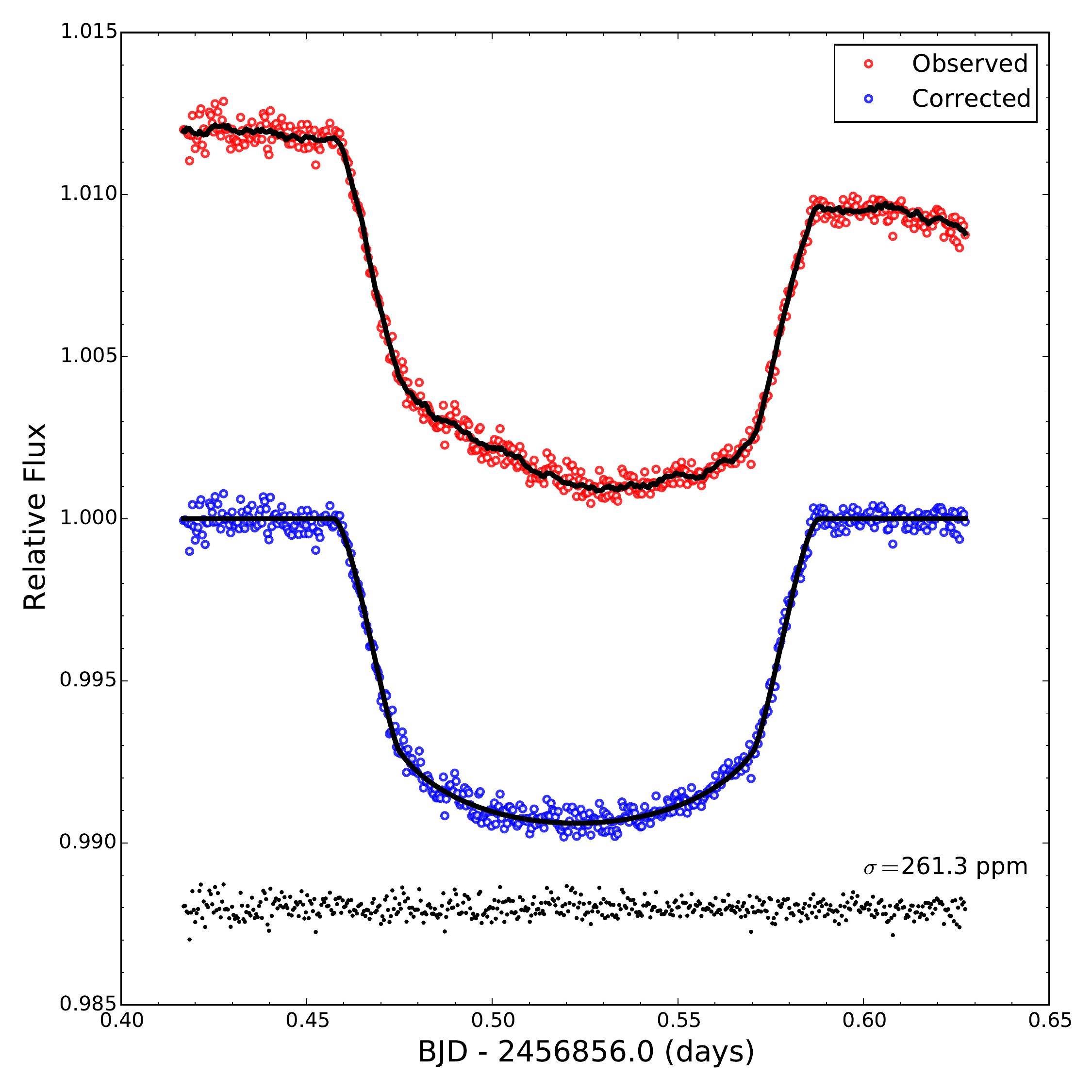}
   \caption{GTC/OSIRIS WASP-48b white light transit curve. The red points present the observed time series and the blue points show the light curve after removing the systematic effects and red noise component. The black line represents the best fit determined using our MCMC analysis. The black dots at the bottom are the residuals of the best fit with an arbitrary offset.}
   \label{Fig:WLCurve}
\end{figure}

The white light curve and best fitted model are shown in Figure \ref{Fig:WLCurve}. The peak-to-peak RMS of the residuals (middle panel of Fig. \ref{Fig:WLCurve}) is 261 ppm. The fitted transit parameters and their corresponding $1\sigma$ uncertainties are listed in Table \ref{table:WL}. The measured planet-to-star radius ratio agrees with a shallower transit than the reported values by \citet{Enoch2011} ($R_\mathrm{p}/R_\mathrm{s} = 0.098 \pm 0.001$) and \citet{Sada2012} ($R_\mathrm{p}/R_\mathrm{s} = 0.0988^{+0.0051}_{-0.0049}$); we speculate that the reason for this discrepancy could be due to spots in the stellar surface.

\begin{table}
  \caption{MCMC results of the white light transit curve of WASP-48b.}
  \label{table:WL}
  \centering
  \begin{tabular}{cc}
    	\hline\hline 
    	Parameter & Value \\
    	\hline 
    	$R_\mathrm{p}/R_\mathrm{s}$ & $0.09330 \pm 0.00088$ \\
	$u_1$ & $0.43583 \pm 0.00183$ \\
	$u_2$ & $0.14877 \pm 0.00466$ \\
	$T_c-2456856$ [days] & $0.522666(234)$ \\
	$a/R_s$ & $4.763 \pm 0.079$ \\
	$i$ [deg] & $82.474 \pm 0.336$ \\
        $e$ & 0 (Fixed) \\
	\hline                                   
  \end{tabular}
\end{table}

\subsection{Transmission spectrum}
\begin{figure*}
   \centering
   \includegraphics[width=\hsize]{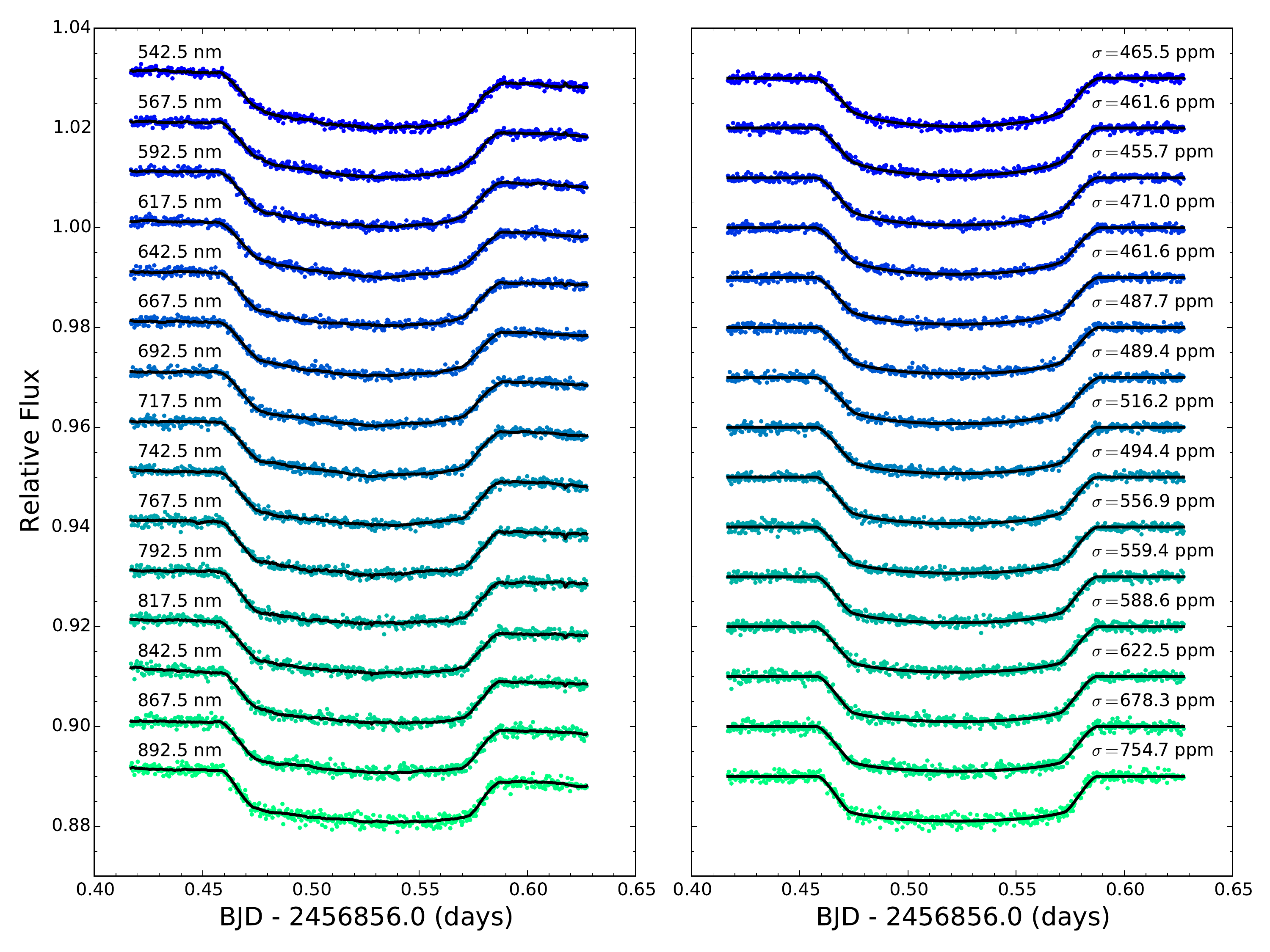}
   \caption{Light curves obtained using the filters of 25 nm of width for WASP48-b. \textit{Left panel}: Observed light curves and best fit (black line). \textit{Right panel}: Transit light curves after removing systematic effects and best fit (black line).}
   \label{Fig:Curves25nm}
 \end{figure*}

The transmission spectrum of WASP-48b was obtained by measuring the change of transit depth across wavelength. Figure \ref{Fig:Curves25nm} shows the time series and best fitted model for the curves made using an integration bin of 25 nm of width. Table \ref{Table:RpRs25nm} presents the measured planet-to-star radius ratio and $1\sigma$ uncertainties for each bin. For all the curves analyzed in this work, the uncertainties in the planet-to-star radius ratio range between $0.8 \times 10^{-3}$ and $1.5\times 10^{-3}$, while the standard deviations of normalized residuals vary between 261 ppm for the white light curve, and 455-755 ppm for spectroscopic light curves. The expected photon noise for the white light curve is 63 ppm and for the spectroscopic light curves it is in the range of 213-373 ppm, meaning that the measured noise level of the white light and spectroscopic curves is close to 4 and 2 times the expected photon noise level, respectively.

\begin{figure*}
   \centering
   \includegraphics[width=\hsize]{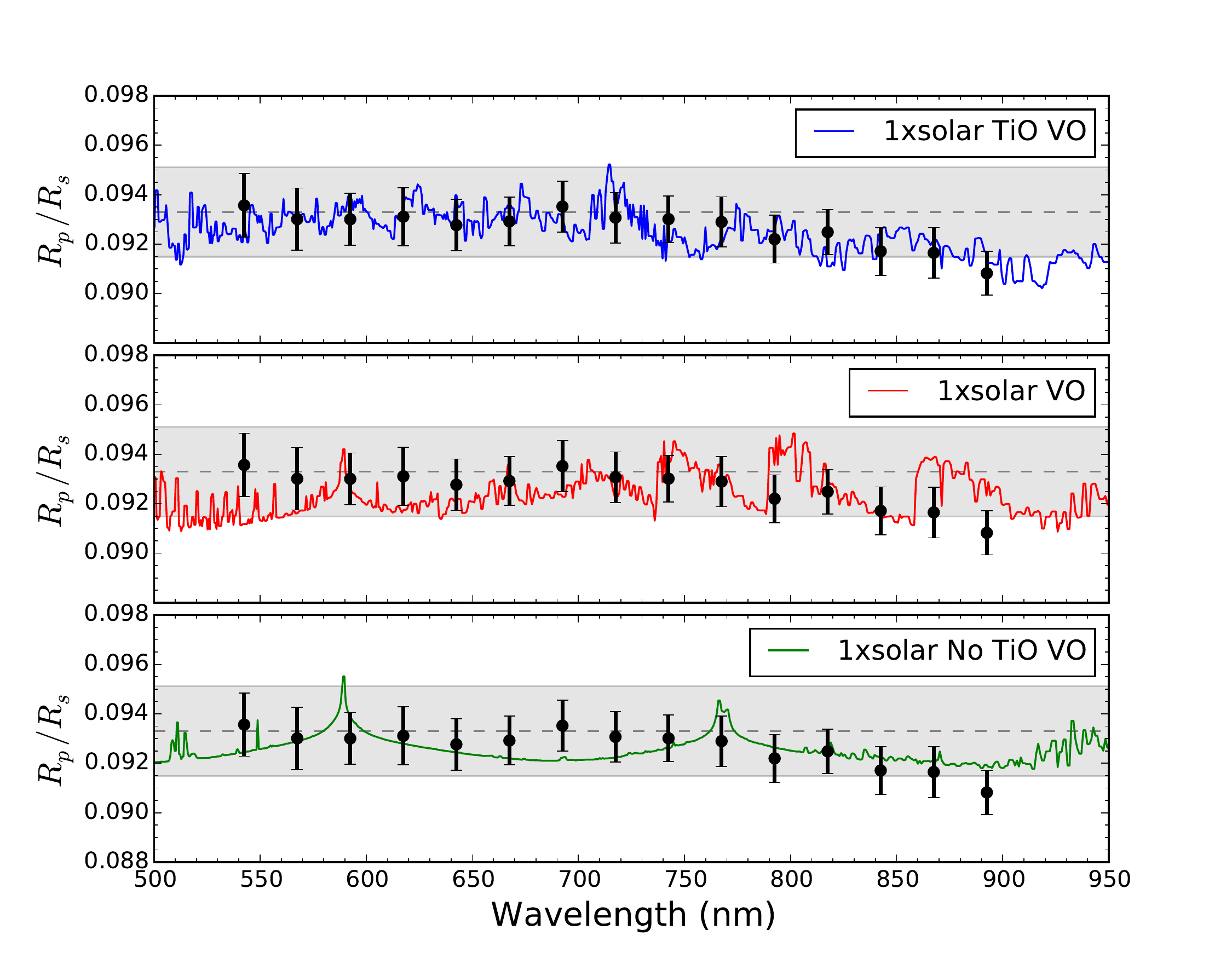}
   \caption{Transmission spectrum of WASP-48b. The solid lines are atmosphere models computed with \textit{Exo-Transmit} for a WASP-48b analog with $T_{eq}=2000$ K, isothermal temperature profile, and Rayleigh scattering by H$_2$ in the blue. The dashed gray line represent the $R_\mathrm{p}/R_\mathrm{s}$ found for the white light curve and the shaded area is $\pm 3$ atmospheric scale heights ($H = 614.9$ km) above and below this level.}
   \label{Fig:TransSpec}
 \end{figure*}

Figure \ref{Fig:TransSpec} presents the transmission spectrum of WASP-48b. In this figure the gray shadowed region corresponds to $\pm 3$ atmospheric scale heights ($H$). One scale height is the distance in which the atmospheric pressure decreases by a factor $e$ (Euler's number) and is given by
\begin{equation}
H = \frac{k_B T_p}{\mu g_p},
\label{Eq:ScaleH}
\end{equation}
where $k_b$ is the Boltzmann constant, $\mu$ is the mean molecular weight, $T_p$ is the temperature of the planet, and $g_p$ is the planet's surface gravity. The atmospheric scale for WASP-48b is $H= 614.9$ km (using $\mu=2.3$ times the proton mass, $T_p=2000$ K, and $ g_p= 11.5$ $m/s^2$ taken from \citealp{Ciceri2015}).

\begin{table}
  \caption{Measured $R_\mathrm{p}/R_\mathrm{s}$ for the 25 nm light curves.}             
  \label{Table:RpRs25nm}      
  \centering                          
  \begin{tabular}{c c}        
    \hline\hline                 
    Center (nm) & $R_\mathrm{p}/R_\mathrm{s}$ \\    
    \hline                        
    542.5 & $0.09357 \pm 0.00128$ \\
    567.5 & $0.09301 \pm 0.00126$ \\
    592.5 & $0.09301 \pm 0.00105$ \\
    617.5 & $0.09311 \pm 0.00117$ \\
    642.5 & $0.09277 \pm 0.00104$ \\
    667.5 & $0.09292 \pm 0.00098$ \\
    692.5 & $0.09352 \pm 0.00103$ \\
    717.5 & $0.09307 \pm 0.00102$ \\
    742.5 & $0.09301 \pm 0.00094$ \\
    767.5 & $0.09290 \pm 0.00102$ \\
    792.5 & $0.09220 \pm 0.00097$ \\
    817.5 & $0.09249 \pm 0.00090$ \\
    842.5 & $0.09171 \pm 0.00097$ \\
    867.5 & $0.09165 \pm 0.00103$ \\
    892.5 & $0.09082 \pm 0.00089$ \\
    \hline                                   
  \end{tabular}
\end{table}

To analyze our transmission spectrum results, we computed theoretical atmosphere models using \textit{Exo-Transmit} (\citealp{Kempton2016}). With this code, we created three cloud-free models with solar composition using an isothermal temperature profile and including Rayleigh scattering, but varying the atmospheric composition. In particular, we computed three models: i) a model atmosphere with TiO and VO, ii) a model atmosphere with VO (no TiO), and iii) a model atmosphere with no TiO and no VO.

\subsection{Refractory element signatures}
The curves made with a bin size of 10 nm were used to put constraints on the presence of sodium in the atmosphere of WASP-48b. This bin size was chosen to cover the Na\,{\sc i} 589.0 and 589.6 nm doublet (\citealp{Murgas2014}). 

The bins near the Na doublet (left and right of the bin centered at the wavelength of the line) were used to compute a weighted mean $R_\mathrm{p}/R_\mathrm{s}$ in order to employ them as a continuum level to see whether there was an extra absorption at the line core. We calculated a continuum level of
\begin{equation}
\left( R_\mathrm{p}/R_\mathrm{s} \right)_{Avg} = 0.09340 \pm 0.00093,
\end{equation}
while the planet-to-star radius ratio at the Na doublet is
\begin{equation}
\left( R_\mathrm{p}/R_\mathrm{s} \right)_{Na} =0.09249 \pm 0.00136.
\end{equation}
The difference between the line and continuum $R_\mathrm{p}/R_\mathrm{s}$ is
\begin{equation}
\Delta \left( R_\mathrm{p}/R_\mathrm{s} \right)_{Avg-Na} = -0.00091 \pm 0.03694,
\end{equation}
meaning that we do not have a statistically significant detection of sodium.

 \begin{figure}
   \centering
   \includegraphics[width=\hsize]{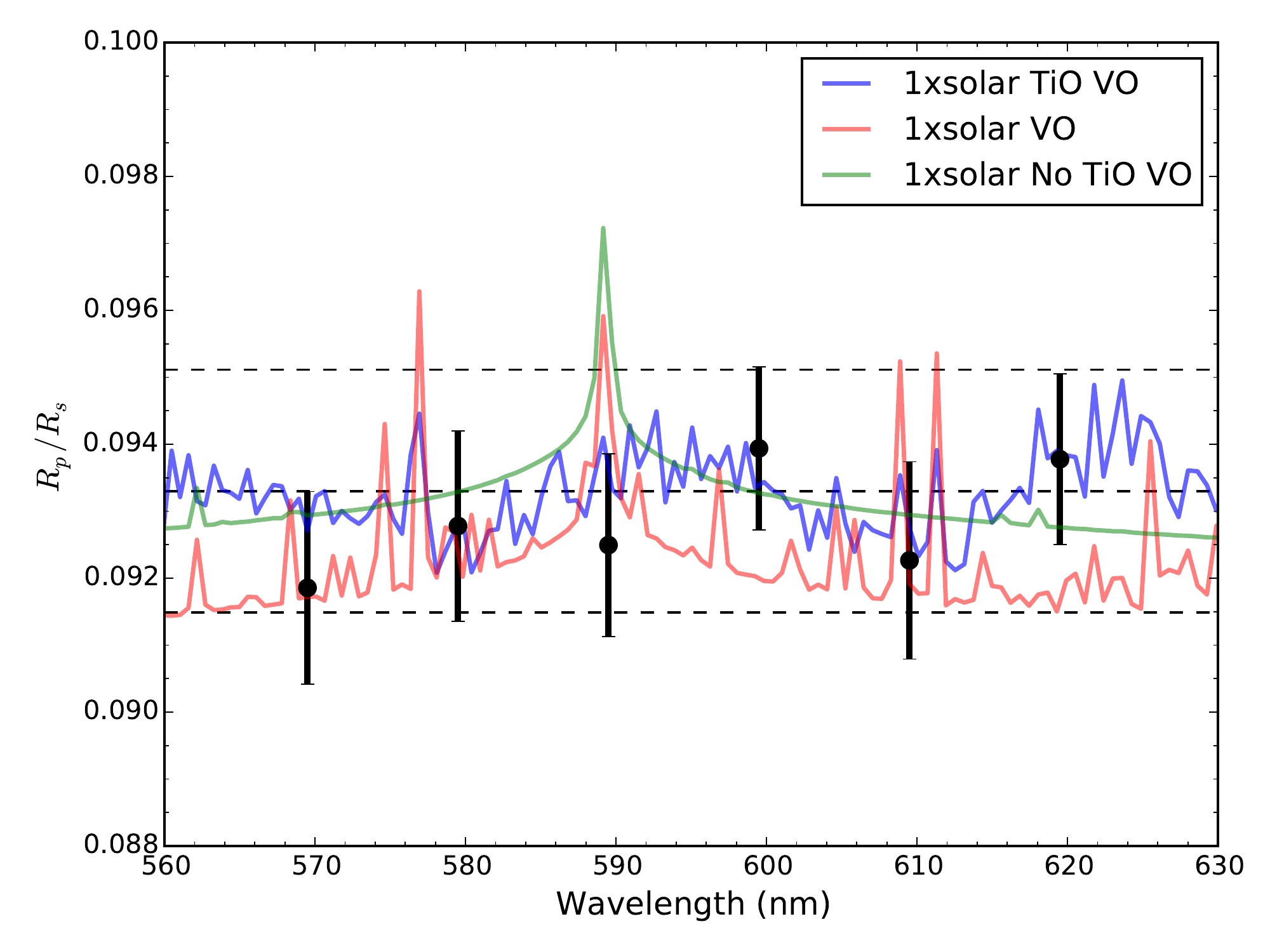}
   \caption{Transmission spectrum of WASP-48b around the Na doublet. The blue, red, and green lines show atmosphere models computed using \textit{Exo-Transmit} for a WASP-48b analog with $T_{eq}=2000$ K, isothermal temperature profile, and Rayleigh scattering by H$_2$ in the blue. The dashed black line represent the $R_\mathrm{p}/R_\mathrm{s}$ found for the white light curve and $\pm 3$ atmospheric scale heights ($H = 614.9$ km) above and below this level.}
   \label{Fig:TransSpec10nm}
 \end{figure}

\begin{table}
  \caption{Measured $R_\mathrm{p}/R_\mathrm{s}$ for the 10 nm light curves.}             
  \label{Table:RpRs10nm}      
  \centering                          
  \begin{tabular}{c c}        
    \hline\hline                 
    Center (nm) & $R_\mathrm{p}/R_\mathrm{s}$ \\    
    \hline                        

    569.5 & $0.09186 \pm 0.00144$ \\
    579.5 & $0.09278 \pm 0.00142$ \\
    589.5 & $0.09249 \pm 0.00136$ \\
    599.5 & $0.09394 \pm 0.00122$ \\
    609.5 & $0.09227 \pm 0.00147$ \\
    619.5 & $0.09378 \pm 0.00128$ \\
        
    \hline                                   
  \end{tabular}
\end{table}

At our resolution the potassium doublet ($\lambda $ 766.48 nm and 769.89 nm) is blended with some atmospheric telluric lines, thus making a detection of an excess in the transit depth difficult due to imperfect telluric correction (\citealp{Parviainen2016}). In this work none of the curves analyzed here presents a significant excess in the transit depth for the bins centered in this line. For the future, the use of a higher resolution grism could help resolve the blended lines and help improve the telluric correction near the K doublet.

\section{Discussion}
\label{sec:discussion}
Several planetary atmosphere models, supported by observations of low mass stars, predict the presence of TiO and VO in hot Jupiters (\citealp{Seager1998}, \citealp{Hubeny2003}, \citealp{Burrows2007}). Moreover, TiO has been proposed as the absorber responsible for the temperature inversion layer observed in some transiting gas giants (\citealp{Fortney2008}). In optical wavelengths, \citet{Desert2008} reported a detection of TiO and VO in the atmosphere of HD 209458b, which is supported by previous albedo measurements for this planet made by \citet{Rowe2006}. \citet{Hoeijmakers2015} searched for the signature of TiO in HD 209458b using cross-correlation techniques applied to high resolution spectra, finding no statistically significant detection of TiO, although they pointed out that their technique depends on the accuracy of TiO line lists used to generate synthetic planet spectrum. A more recently discovered exoplanet with tentative detection of TiO and VO in the optical is the highly inflated WASP-127b (\citealp{Lam2017}, \citealp{Palle2017}), although more conclusive studies need to be made. The detection of TiO and VO molecules in the optical is difficult because their features are thought to be relatively broad and of low amplitude, in addition the presence of hazes in the planetary atmospheres can affect the strength of the features (\citealp{Pont2013}, \citealp{Kreidberg2014}, \citealp{Burrows2014} and references therein).

\citet{ORourke2014} measured the secondary transit of WASP-48b in the infrared ($H$, $K_s$, 3.6 $\mu$m, and 4.5 $\mu$m bands). By comparing their data to the models of \citet{Fortney2008} and \citet{Burrows2008}, they deduced that WASP-48b has a weak or absent temperature inversion and moderate day-night energy circulation. The model that presented the best fit was one without TiO, but models with a more efficient day-night circulation and including TiO could also provide a reasonably close match to the data. 

Using $\chi^2$ statistics to fit the models to the observed transmission spectrum, we found that the preferred atmosphere model was the one including TiO and VO (top panel in Fig. \ref{Fig:TransSpec}) with a reduced chi square of $\chi_r^2 = 0.32$ (15 degrees of freedom); for comparison a straight line model gave $\chi_r^2 = 0.59$. For the full observed wavelength range, the model including TiO and VO presented a lower $\chi_r^2$ value, although it is not statistically significant to be differentiated from a featureless spectrum, i.e, a flat line. The model with only VO and the model without TiO and VO, had a reduced chi square of $\chi_r^2 = 0.85$ and $\chi_r^2 = 2.27$, respectively. 

After 710 nm the large-scale structure of the TiO bands has a decreasing slope (e.g., \citealp{Sharp2007}, \citealp{Fortney2010}), which coincides with the change in slope in the observed transmission spectrum. We divided the transmission spectrum in two regions: bins with $\lambda < 710 $ nm and bins with $\lambda \geq 710$ nm, and we performed a simple linear fit for both regions. The measured slope in the region with $\lambda < 710 $ was consistent with a flat line (slope value of $(-0.320 \pm 8.545)\times 10^{-6}$), while the bins with $\lambda \geq 710$ nm presented a fitted slope value of $(-12.57 \pm 5.94)\times 10^{-6}$. Although our data seems to show two regions with different slopes, the spectral resolution and uncertainties of the observed transmission spectrum of WASP-48b does not allow us to claim a detection of TiO and VO.

\section{Conclusions}
\label{sec:conclusions}
We present here an analysis of a primary transit of WASP-48b taken with GTC/OSIRIS instrument. Using a time series of long-slit spectra of WASP-48 and one reference star, we created several color light curves in order to explore the change in transit depth across wavelength. 

In this work we analyzed 1 white light curve (wavelength range 530-905 nm), 15 curves of 25 nm in width (wavelength range 530-905 nm), and 6 curves of 10 nm (wavelength range 564.5-624.5 nm). All the curves were fitted using a Bayesian MCMC procedure using a transit model with a quadratic limb darkening law and reproducing the systematic effects present in the curves. The uncertainties reported in this work take into account time correlated noise of unknown origin (red noise) computed using Gaussian processes. 

We report a flat, featureless transmission spectrum confirming previous broadband observations. The obtained transmission spectrum agrees with the expected cloud-free theoretical atmosphere model that includes the presence of TiO and VO, although evidence of these molecules in the atmosphere of WASP-48b is not statistically significant enough to claim a detection. Exploring the wavelength region near the Na\,{\sc i} doublet ($\lambda$ 589.0 and 589.6 nm), we found no statistically significant detection of this element. 
  
\begin{acknowledgements}
 Based on observations made with the Gran Telescopio Canarias (GTC), installed in the Spanish Observatorio del Roque de los Muchachos of the Instituto de Astrof\'isica de Canarias, in the island of La Palma. All the figures presented here were made using Matplotlib (\citealp{Hunter2007}).
\end{acknowledgements}

%
%

\bibliographystyle{aa}
\bibliography{biblio}

\end{document}